\documentclass[11pt, letterpaper]{article}
\usepackage{amsfonts,amsmath,amsopn,amssymb,amsthm,bbm,latexsym,mathrsfs,pstricks,verbatim,color}

\font\small=cmr8

\font\tenmsy=msbm10
\font\sevenmsy=msbm10 at 7pt
\font\fivemsy=msbm10 at 5pt
\newfam\msyfam 
\textfont\msyfam=\tenmsy
\scriptfont\msyfam=\sevenmsy
\scriptscriptfont\msyfam=\fivemsy

\let\e\epsilon
\let\s\sigma

\let\l\left
\let\r\right

\def\z{{\cal Z}}
\let\lf\lfloor
\let\rf\rfloor

\def\y{{\infty}}

\let\Rw\Rightarrow
\def\l{{\left}}
\def\r{{\right}}

\def\rw{\rightarrow}

\def\frac#1#2{{#1 \over #2}}

\def\w{{\tilde w}}

\let\ka\kappa

\font\small=cmr8

\begin{document}

\let\Rw\Rightarrow
\let\rw\rightarrow
\let\l\left
\let\r\right
\let\s\sigma
\let\ka\kappa
\let\de\delta

\def\M{{\cal M}}
\def\SM{{\cal SM}}

 \let\g\gamma
 
\vskip18pt

\title{\vskip60pt {\bf Particles in  RSOS paths }}

\vskip18pt

\smallskip
\author{ \bf{P. Jacob and P.
Mathieu}\thanks{patrick.jacob@hotmail.com,
pmathieu@phy.ulaval.ca.} \\ 
\\
D\'epartement de physique, de g\'enie physique et d'optique,\\
Universit\'e Laval,
Qu\'ebec, Canada, G1K 7P4.}

\vskip .2in
\bigskip

\maketitle


\vskip0.3cm
\centerline{{\bf ABSTRACT}}
\vskip18pt

We introduce a new representation of the paths  of the Forrester-Baxter RSOS models which represents the states of the irreducible modules of  the minimal models $\M(p',p)$.  This representation   is obtained by 
transforming the  RSOS paths, for the cases $p\geq 2p'-1$,
to new paths 
 for which  horizontal edges are allowed at certain heights.
These new paths are much simpler in that 
 their  weight 
 is nothing but the sum of the position of the peaks. This description paves the way for the interpretation of the RSOS paths in terms of fermi-type charged particles out of which the fermionic characters could be obtained constructively.
The derivation of the  fermionic character for $p'=2$ and $p=kp'\pm 1$
is outlined.
Finally, the particles of the RSOS paths are put in relation with the kinks and the breathers  of the restricted sine-Gordon model.





\section{Introduction}

All the minimal models $\M(p',p)$ (where $p$ and $p'$ are coprime and  $p>p'$) have a  restricted-solid-on-solid (RSOS) path description. It means that the  states in irreducible modules can be represented by the infinite-length limit of  lattice paths,
 with the initial and final points characterizing the module. 
 The link between the generating function of paths (or equivalently, configuration sums, since a path 
 forms the contour of a RSOS configuration) and characters, discovered in \cite{Kyoto},
relies on  the corner-transfer matrix solution \cite{Ba} of the statistical model in the so-called regime III \cite{ABF,FB}.
 
 The RSOS paths with parameters $(p',p)$ -- referred to as the RSOS$(p',p)$ paths -- provide thus a combinatorial description
 of the set of states of the $\M(p',p)$ model \cite{Mel}. 
 

The fermionic character  \cite{KKMMa} of all irreducible modules can be  derived from this  path description \cite{FLPW}. 
However, the method used in these latter references does not make clear the particle structure of the generic  RSOS$(p',p)$ paths that  underlies the fermionic character. This prevents a direct constructive (as opposed to a recursive) approach to these fermionic forms.
%
 This is to be  contrasted with the analysis of \cite{OleJS} for the special (and much simpler) class of unitary models ($p'=p-1)$. There, a RSOS path is interpreted 
 as a closely packed configuration of one-dimensional charged fermionic-type particles with specific exclusion rules. The character  is obtained constructively and expressed as a multiple-sum over the number of  particles of different  type.
 
There are two other classes of minimal models that do have a path representation with a neat particle description: the Yang-Lee sequence $\M(2,2k+1)$ \cite{W97} (following their analysis in  \cite{FNO}) and the $\M(k+1,2k+3)$ \cite{JSTAT} ones.\footnote{T. Welsh reports
that he has found a bijection relating the path description of the $\M(k+1,2k+3)$ models found in  \cite{JSTAT} to the standard RSOS paths (private communication).}

 Here, we present a correspondence between RSOS$(p',p)$ paths, for $p\geq 2p'-1$, with a new type of paths, which we call generalized Bressoud paths, to be denoted as B$(p',p)$. 
 The original Bressoud paths \cite{BreL}  provide still another path description of the $\M(2,2k+1)$ models (combining the results of \cite{FNO,Bu,BreL}). In addition to North-East (NE) and South-East (SE) edges -- out of which RSOS paths are composed -- ,  horizontal (H) edges are allowed on the $x$-axis.
For the generalized Bressoud paths B$(p',p)$,
H edges are allowed at specific heights, determined by the ratio $p/p'$. 
The crux of the matter with this transformation is that the weight of a B$(p',p)$ path is simply the horizontal position of its peaks plus  half the position of its `half-peaks'. This is a key step for the reinterpretation  of the collection of paths in terms of a gas of fermi-type charged particles.

But what is meant by a {\it particle interpretation of a path}?
If the paths can be decomposed into a number of building blocks whose dynamics (by which we refer to the block displacements and the interpenetration patterns of the  blocks of different types) generate all possible paths, then we say that we have a particle interpretation, the particles being identified with the basic blocks. 
Typically the particles are elementary  triangles, as in the unitary case or the $\M(2,2k+1)$ models, or generalized triangles incorporating some H edges, as observed in the new paths obtained here but also for the paths pertaining to the superconformal models \cite{JMsusy}.  The charge  qualifier in the expression `fermi-type charged particles' refers to a label that distinguishes the different particles, usually related to the triangle width. Their ferminionc nature reflects the impenetrability of identical particles.

Our main point here is that the transformation of RSOS paths to generalized Bressoud paths is instrumental for reading off a natural set of particles. 
Once we have a particle description of a class of paths, it is usually possible to obtain their fermionic character by
a direct constructive method. This is illustrated here  for the general $\M(p',kp'\pm 1)$ models.


To which extent is this particle description physical? Granting its uniqueness, 
we expect these particles to be related to the (yet to be fully worked out) quasi-particles underlying a representation of the minimal-model irreducible modules. Such a representation would provide a substitute to the usual representation theory based on the Virasoro algebra and its singular vectors. 
But more immediately,  this should match the particle description of the off-critical theory.
The RSOS model in regime III provides an off-critical lattice description of the minimal models perturbed by the $\phi_{1,3}$ field \cite{Huse}, a perturbation that preserves integrability \cite{Zam}. 
The spectrum of RSOS particles should thus  be compared to that of these integrable models. 
This is addressed  in section 5 where the RSOS particles are related to the kinks and the breathers of the restricted sine-Gordon model \cite{LeC}.

\section{RSOS$(p',p)$ paths}

A configuration pertaining to the  RSOS realization of the finitized $\M(p',p)$  minimal models (with $p>p'$) is described by a sequence of values of the height variables $\ell_i\in \{1,2,\cdots, p-1\}$. The height index  is bounded by  $0\leq i\leq L$.  Adjacent heights are subject to the admissibility condition: $|\ell_i-\ell_{i+1}|=1$. 
A path 
is the contour of a configuration. It is thus a sequence of NE or SE edges joining the adjacent vertices $(i,\ell_i)$ and $(i+1,\ell_{i+1})$ of the configuration. 
Each path is specified by particular boundary conditions: the values of $\ell_0$ and those of $\ell_{L-1}$ and $\ell_L$ (the last two specify the RSOS ground state on which the configuration is built). 
For definiteness, we will choose $\ell_{L-1}=\ell_L+1$ so that  the paths all terminate with a SE edge.

The weight of a path is the sum of  the weight of all its non-extremal vertices:
\begin{equation}\label{wei}
\w=\sum_{i=1}^{L-1} \w_i.
\end{equation}
 With each type of vertex specified by the triplet $(\ell_{i-1},\ell_i, \ell_{i+1})$,
 we have -- in regime III -- \cite{FB}:
\begin{align}\label{wFB}
(h\mp1,h,h\pm1): &\quad \w_i = \frac{i}2,\nonumber\\
(h,h\mp1,h): &\quad \w_i =\pm i  \l\lf \frac{h(p-p')}{p} \r\rf,
\end{align} 
with $\lf u\rf $ standing  for the largest integer smaller than $u$.

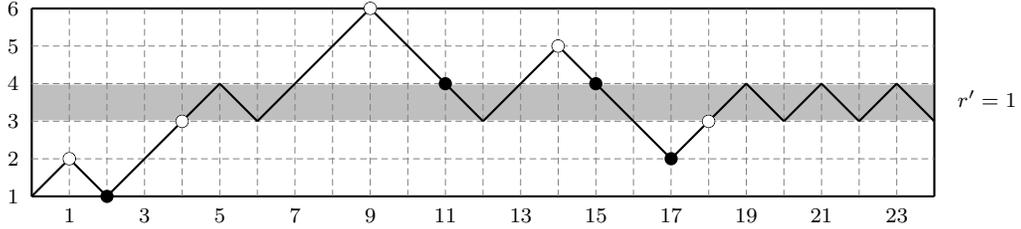
\begin{figure}[ht]
\caption{{\footnotesize A typical path pertaining to the vacuum module $(s=r=1)$ for the $\M(2,7) $ model (the identification of a module in terms of the end points is explained at the end of the section, in eq. (\ref{kacla})). The weight $\w$ of this path obtained using the expression (\ref{wFB}) is $\w=131/2$. The corresponding ground-state configuration is the path which starts with two NE edges and is followed by a sequence of 11 pairs of (NE,SE) edges. Its weight is $47/2$. The relative weight of the pictured path is thus $\Delta \w =  \w-\w_{{\rm gsc}}=42$. 
The dots indicate the scoring vertices which refer to the alternative weight description of the path explained below in the main text. The white dots have weights $u_i$ and the black ones have weights $v_i$, these being defined in eq. (\ref{defuv}).  The scoring vertices have respective weight  $0,1,1,2,7,5,9,9,8$, so that the weight $w$ of the path is $w=42$, illustrating the identity $\Delta \w=w$.}}
\vskip-1.4cm
\label{fig1}
\begin{center}
\begin{pspicture}(0,0)(13.0,4.5)



{
\psset{linestyle=none}
\psset{fillstyle=solid}
\psset{fillcolor=lightgray}
\psframe(0.5,1.5)(12.5, 2.0)
}


\psset{linestyle=solid}
\psline{-}(0.5,0.5)(12.5,0.5)
\psline{-}(0.5,3.0)(12.5,3.0)

\psline{-}(0.5,0.5)(0.5,3.0)
\psline{-}(12.5,0.5)(12.5,3.0)

{
\psset{linewidth=0.25pt,linestyle=dashed, dash=2.5pt 1.5pt,linecolor=gray}

\psline{-}(0.5,1.0)(12.5,1.0)
\psline{-}(0.5,1.5)(12.5,1.5)
\psline{-}(0.5,2.0)(12.5,2.0)
\psline{-}(0.5,2.5)(12.5,2.5)

\psline{-}(1.0,0.5)(1.0,3.0) \psline{-}(1.5,0.5)(1.5,3.0) 
\psline{-}(2.0,0.5)(2.0,3.0) \psline{-}(2.5,0.5)(2.5,3.0) \psline{-}(3.0,0.5)(3.0,3.0) 
\psline{-}(3.5,0.5)(3.5,3.0) \psline{-}(4.0,0.5)(4.0,3.0) \psline{-}(4.5,0.5)(4.5,3.0) 
\psline{-}(5.0,0.5)(5.0,3.0) \psline{-}(5.5,0.5)(5.5,3.0) \psline{-}(6.0,0.5)(6.0,3.0) 
\psline{-}(6.5,0.5)(6.5,3.0) \psline{-}(7.0,0.5)(7.0,3.0) \psline{-}(7.5,0.5)(7.5,3.0) 
\psline{-}(8.0,0.5)(8.0,3.0) \psline{-}(8.5,0.5)(8.5,3.0) \psline{-}(9.0,0.5)(9.0,3.0) 
\psline{-}(9.5,0.5)(9.5,3.0) \psline{-}(10.0,0.5)(10.0,3.0) \psline{-}(10.5,0.5)(10.5,3.0) 
\psline{-}(11.0,0.5)(11.0,3.0) \psline{-}(11.5,0.5)(11.5,3.0) \psline{-}(12.0,0.5)(12.0,3.0) 
}



{\scriptsize
\rput(1.0,0.25){{$1$}} \rput(2.0,0.25){{$3$}}
\rput(3.0,0.25){{$5$}} \rput(4.0,0.25){{$7$}}
\rput(5.0,0.25){{$9$}} \rput(6.0,0.25){{$11$}}
\rput(7.0,0.25){{$13$}} \rput(8.0,0.25){{$15$}}
\rput(9.0,0.25){{$17$}} \rput(10.0,0.25){{$19$}}
\rput(11.0,0.25){{$21$}} \rput(12.0,0.25){{$23$}}
\rput(13.2,1.8){{$r'=1$}} 
}



\rput(0.25,0.5){{\scriptsize $1$}}
\rput(0.25,1.0){{\scriptsize $2$}} \rput(0.25,1.5){{\scriptsize $3$}}
\rput(0.25,2.0){{\scriptsize $4$}} \rput(0.25,2.5){{\scriptsize $5$}}
\rput(0.25,3.0){{\scriptsize $6$}}



\psset{linestyle=solid}

\psline{-}(0.5,0.5)(1.0,1.0) \psline{-}(1.0,1.0)(1.5,0.5)
\psline{-}(1.5,0.5)(2.0,1.0) \psline{-}(2.0,1.0)(2.5,1.5)
\psline{-}(2.5,1.5)(3.0,2.0) \psline{-}(3.0,2.0)(3.5,1.5)
\psline{-}(3.5,1.5)(4.0,2.0) \psline{-}(4.0,2.0)(4.5,2.5)
\psline{-}(4.5,2.5)(5.0,3.0) \psline{-}(5.0,3.0)(5.5,2.5)
\psline{-}(5.5,2.5)(6.0,2.0) \psline{-}(6.0,2.0)(6.5,1.5)
\psline{-}(6.5,1.5)(7.0,2.0) \psline{-}(7.0,2.0)(7.5,2.5)
\psline{-}(7.5,2.5)(8.0,2.0) \psline{-}(8.0,2.0)(8.5,1.5)
\psline{-}(8.5,1.5)(9.0,1.0)\psline{-}(9.0,1.0)(9.5,1.5)
\psline{-}(9.5,1.5)(10.0,2.0)\psline{-}(10.0,2.0)(10.5,1.5)
\psline{-}(10.5,1.5)(11.0,2.0)\psline{-}(11.0,2.0)(11.5,1.5)
\psline{-}(11.5,1.5)(12.0,2.0)\psline{-}(12.0,2.0)(12.5,1.5)


\psset{fillcolor=white}
\psset{dotsize=5pt}\psset{dotstyle=o}
\psdots(1,1)(5,3)(7.5,2.5)(2.5,1.5)(9.5,1.5)
\psset{dotsize=5pt}\psset{dotstyle=*}
\psdots(6.0,2.0)(8.0,2.0)(1.5,0.5)(9,1)

\end{pspicture}
\end{center}
\end{figure}

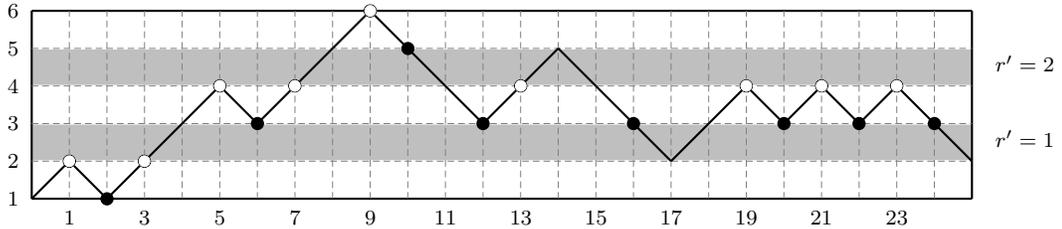
\begin{figure}[ht]
\caption{{\footnotesize The path of Fig.\ref{fig1}  for  the  $\M(3,7)$ model  augmented with a SE edge, with its scoring vertices indicated.  
It again describes a state in the vacuum module $s=r=1$.  The corresponding ground state configuration is the path with a single NE edge followed by a zigzag pattern in the first gray band. The relative weight of the pictured path is found to be $\Delta \w=92$.}}
\vskip-1.4cm
\label{fig2}
\begin{center}
\begin{pspicture}(0,0)(13.0,4.5)




{
\psset{linestyle=none}
\psset{fillstyle=solid}
\psset{fillcolor=lightgray}
\psframe(0.5,1.0)(13, 1.5)
\psframe(0.5,2.0)(13, 2.5)
}

\psset{linestyle=solid}
\psline{-}(0.5,0.5)(13,0.5)
\psline{-}(0.5,3.0)(13,3.0)

\psline{-}(0.5,0.5)(0.5,3.0)
\psline{-}(13,0.5)(13,3.0)

{
\psset{linewidth=0.25pt,linestyle=dashed, dash=2.5pt 1.5pt,linecolor=gray}

\psline{-}(0.5,1.0)(13,1.0)
\psline{-}(0.5,1.5)(13,1.5)
\psline{-}(0.5,2.0)(13,2.0)
\psline{-}(0.5,2.5)(13,2.5)

\psline{-}(1.0,0.5)(1.0,3.0) \psline{-}(1.5,0.5)(1.5,3.0) 
\psline{-}(2.0,0.5)(2.0,3.0) \psline{-}(2.5,0.5)(2.5,3.0) \psline{-}(3.0,0.5)(3.0,3.0) 
\psline{-}(3.5,0.5)(3.5,3.0) \psline{-}(4.0,0.5)(4.0,3.0) \psline{-}(4.5,0.5)(4.5,3.0) 
\psline{-}(5.0,0.5)(5.0,3.0) \psline{-}(5.5,0.5)(5.5,3.0) \psline{-}(6.0,0.5)(6.0,3.0) 
\psline{-}(6.5,0.5)(6.5,3.0) \psline{-}(7.0,0.5)(7.0,3.0) \psline{-}(7.5,0.5)(7.5,3.0) 
\psline{-}(8.0,0.5)(8.0,3.0) \psline{-}(8.5,0.5)(8.5,3.0) \psline{-}(9.0,0.5)(9.0,3.0) 
\psline{-}(9.5,0.5)(9.5,3.0) \psline{-}(10.0,0.5)(10.0,3.0) \psline{-}(10.5,0.5)(10.5,3.0) 

\psline{-}(11.0,0.5)(11.0,3.0) \psline{-}(11.5,0.5)(11.5,3.0) \psline{-}(12.0,0.5)(12.0,3.0) 
\psline{-}(12.5,0.5)(12.5,3.0)
}



\rput(13.7,2.3){{\scriptsize$r'=2$}} 
\rput(13.7,1.3){{\scriptsize$r'=1$}} 
\rput(1.0,0.25){{\scriptsize $1$}} \rput(2.0,0.25){{\scriptsize $3$}}
\rput(3.0,0.25){{\scriptsize $5$}} \rput(4.0,0.25){{\scriptsize $7$}}
\rput(5.0,0.25){{\scriptsize $9$}} \rput(6.0,0.25){{\scriptsize $11$}}
\rput(7.0,0.25){{\scriptsize $13$}} \rput(8.0,0.25){{\scriptsize $15$}}
\rput(9.0,0.25){{\scriptsize $17$}} \rput(10.0,0.25){{\scriptsize $19$}}
\rput(11.0,0.25){{\scriptsize $21$}} \rput(12.0,0.25){{\scriptsize $23$}}

\rput(0.25,0.5){{\scriptsize $1$}}
\rput(0.25,1.0){{\scriptsize $2$}} \rput(0.25,1.5){{\scriptsize $3$}}
\rput(0.25,2.0){{\scriptsize $4$}} \rput(0.25,2.5){{\scriptsize $5$}}
\rput(0.25,3.0){{\scriptsize $6$}}



\psset{linestyle=solid}

\psline{-}(0.5,0.5)(1.0,1.0) \psline{-}(1.0,1.0)(1.5,0.5)
\psline{-}(1.5,0.5)(2.0,1.0) \psline{-}(2.0,1.0)(2.5,1.5)
\psline{-}(2.5,1.5)(3.0,2.0) \psline{-}(3.0,2.0)(3.5,1.5)
\psline{-}(3.5,1.5)(4.0,2.0) \psline{-}(4.0,2.0)(4.5,2.5)
\psline{-}(4.5,2.5)(5.0,3.0) \psline{-}(5.0,3.0)(5.5,2.5)
\psline{-}(5.5,2.5)(6.0,2.0) \psline{-}(6.0,2.0)(6.5,1.5)
\psline{-}(6.5,1.5)(7.0,2.0) \psline{-}(7.0,2.0)(7.5,2.5)
\psline{-}(7.5,2.5)(8.0,2.0) \psline{-}(8.0,2.0)(8.5,1.5)
\psline{-}(8.5,1.5)(9.0,1.0)\psline{-}(9.0,1.0)(9.5,1.5)
\psline{-}(9.5,1.5)(10.0,2.0)\psline{-}(10.0,2.0)(10.5,1.5)
\psline{-}(10.5,1.5)(11.0,2.0)\psline{-}(11.0,2.0)(11.5,1.5)

\psline{-}(11.5,1.5)(12.0,2.0)\psline{-}(12.0,2.0)(12.5,1.5)
\psline{-}(12.5,1.5)(13.0,1.0)


\psset{fillcolor=white}
\psset{dotsize=5pt}\psset{dotstyle=o}
\psdots(1,1)(2,1)(3,2)(4,2)(5,3)(7,2)(10,2)(11,2)(12,2)
\psset{dotsize=5pt}\psset{dotstyle=*}
\psdots(1.5,0.5)(3.5,1.5)(5.5,2.5)(6.5,1.5)(8.5,1.5)(10.5,1.5)(11.5,1.5)(12.5,1.5)

\end{pspicture}
\end{center}
\end{figure}

\begin{figure}[ht]
\caption{{\footnotesize The path of Fig.\ref{fig1}  augmented with 2 SE edges, in the context of  the $\M(4,7)$ model. It  represents a state in the vacuum module. If the path were cut at $L=24$, it would instead correspond to  a state in the module with $s=1,\, r=2$.}}
\vskip-1.4cm
\label{fig3}
\begin{center}
\begin{pspicture}(1,0)(13.0,4.5)




{
\psset{linestyle=none}
\psset{fillstyle=solid}
\psset{fillcolor=lightgray}
\psframe(0.5,0.5)(13.5, 1.0)
\psframe(0.5,1.5)(13.5, 2.0)
\psframe(0.5,2.5)(13.5, 3.0)
}

\psset{linestyle=solid}
\psline{-}(0.5,0.5)(13.5,0.5)
\psline{-}(0.5,3.0)(13.5,3.0)

\psline{-}(0.5,0.5)(0.5,3.0)
\psline{-}(13.5,0.5)(13.5,3.0)

{
\psset{linewidth=0.25pt,linestyle=dashed, dash=2.5pt 1.5pt,linecolor=gray}

\psline{-}(0.5,1.0)(13.5,1.0)
\psline{-}(0.5,1.5)(13.5,1.5)
\psline{-}(0.5,2.0)(13.5,2.0)
\psline{-}(0.5,2.5)(13.5,2.5)

\psline{-}(1.0,0.5)(1.0,3.0) \psline{-}(1.5,0.5)(1.5,3.0) 
\psline{-}(2.0,0.5)(2.0,3.0) \psline{-}(2.5,0.5)(2.5,3.0) \psline{-}(3.0,0.5)(3.0,3.0) 
\psline{-}(3.5,0.5)(3.5,3.0) \psline{-}(4.0,0.5)(4.0,3.0) \psline{-}(4.5,0.5)(4.5,3.0) 
\psline{-}(5.0,0.5)(5.0,3.0) \psline{-}(5.5,0.5)(5.5,3.0) \psline{-}(6.0,0.5)(6.0,3.0) 
\psline{-}(6.5,0.5)(6.5,3.0) \psline{-}(7.0,0.5)(7.0,3.0) \psline{-}(7.5,0.5)(7.5,3.0) 
\psline{-}(8.0,0.5)(8.0,3.0) \psline{-}(8.5,0.5)(8.5,3.0) \psline{-}(9.0,0.5)(9.0,3.0) 
\psline{-}(9.5,0.5)(9.5,3.0) \psline{-}(10.0,0.5)(10.0,3.0) \psline{-}(10.5,0.5)(10.5,3.0) 
\psline{-}(11.0,0.5)(11.0,3.0) \psline{-}(11.5,0.5)(11.5,3.0) \psline{-}(12.0,0.5)(12.0,3.0) 
\psline{-}(12.5,0.5)(12.5,3.0) \psline{-}(13.0,0.5)(13.0,3.0) 
}





\rput(14.2,2.8){{\scriptsize$r'=3$}} 
\rput(14.2,1.8){{\scriptsize$r'=2$}} 
\rput(14.2,0.8){{\scriptsize$r'=1$}} 
\rput(1.0,0.25){{\scriptsize $1$}} \rput(2.0,0.25){{\scriptsize $3$}}
\rput(3.0,0.25){{\scriptsize $5$}} \rput(4.0,0.25){{\scriptsize $7$}}
\rput(5.0,0.25){{\scriptsize $9$}} \rput(6.0,0.25){{\scriptsize $11$}}
\rput(7.0,0.25){{\scriptsize $13$}} \rput(8.0,0.25){{\scriptsize $15$}}
\rput(9.0,0.25){{\scriptsize $17$}} \rput(10.0,0.25){{\scriptsize $19$}}
\rput(11.0,0.25){{\scriptsize $21$}} \rput(12.0,0.25){{\scriptsize $23$}}
\rput(13.0,0.25){{\scriptsize $25$}}

\rput(0.25,0.5){{\scriptsize $1$}}
\rput(0.25,1.0){{\scriptsize $2$}} \rput(0.25,1.5){{\scriptsize $3$}}
\rput(0.25,2.0){{\scriptsize $4$}} \rput(0.25,2.5){{\scriptsize $5$}}
\rput(0.25,3.0){{\scriptsize $6$}}



\psset{linestyle=solid}

\psline{-}(0.5,0.5)(1.0,1.0) \psline{-}(1.0,1.0)(1.5,0.5)
\psline{-}(1.5,0.5)(2.0,1.0) \psline{-}(2.0,1.0)(2.5,1.5)
\psline{-}(2.5,1.5)(3.0,2.0) \psline{-}(3.0,2.0)(3.5,1.5)
\psline{-}(3.5,1.5)(4.0,2.0) \psline{-}(4.0,2.0)(4.5,2.5)
\psline{-}(4.5,2.5)(5.0,3.0) \psline{-}(5.0,3.0)(5.5,2.5)
\psline{-}(5.5,2.5)(6.0,2.0) \psline{-}(6.0,2.0)(6.5,1.5)
\psline{-}(6.5,1.5)(7.0,2.0) \psline{-}(7.0,2.0)(7.5,2.5)
\psline{-}(7.5,2.5)(8.0,2.0) \psline{-}(8.0,2.0)(8.5,1.5)
\psline{-}(8.5,1.5)(9.0,1.0)\psline{-}(9.0,1.0)(9.5,1.5)
\psline{-}(9.5,1.5)(10.0,2.0)\psline{-}(10.0,2.0)(10.5,1.5)
\psline{-}(10.5,1.5)(11.0,2.0)\psline{-}(11.0,2.0)(11.5,1.5)
\psline{-}(11.5,1.5)(12.0,2.0)\psline{-}(12.0,2.0)(13.5,.5)


\psset{fillcolor=white}
\psset{dotsize=5pt}\psset{dotstyle=o}
\psdots(2.5,1.5)(4.5,2.5)(7.5,2.5)(9.5,1.5)
\psset{dotsize=5pt}\psset{dotstyle=*}
\psdots(6.0,2.)(8,2.)(9,1)(13,1)

\end{pspicture}
\end{center}
\end{figure}
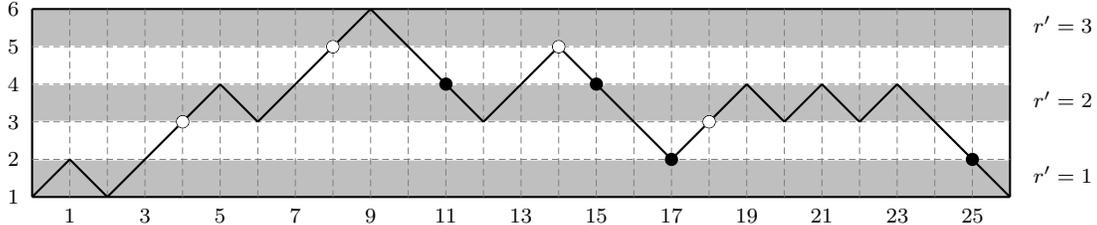

The rectangle delimiting the RSOS$(p',p)$ paths can be made $p'$-dependent looking by coloring in gray the $p'-1$ strips between those heights $h$ and $h+1$ for which \cite{FLPW}:
\begin{equation}
\l \lf \frac{hp'}{p}\r\rf =  \l \lf \frac{(h+1)p'}{p}\r\rf -1.
\end{equation}
It is easily verified that the values of $h$ which satisfy this condition are the following:
\begin{equation}\label{hr}
h_{r'}=  \l \lf \frac{r'p}{p'}\r\rf \qquad \text{for} \qquad 1\leq r'\leq p'-1 .
\end{equation}
The upper bound on $r'$ follows from the condition $h\leq p-1$.
The band structure is symmetric with respect to the up-down
reversal.
For unitary models, $p=p'+1$, all the bands are gray. In the other extreme case, for the $\M(2,p)$ models, there is a single gray band, right in the middle.
The band structure is illustrated in Figs \ref{fig1}-\ref{fig3} where basically the same path is drawn but within rectangles pertaining to the $\M(p',7)$ models with $p'=2,3$ and 4.


For reasons to be explained shortly, our analysis is restricted to those models for which the gray bands are isolated. This is so when
\begin{equation}\label{bou}
p\geq 2p'-1.
\end{equation}
For $p=7$, the three cases illustrated are precisely those that satisfy this bound.

With the  boundary conditions specified, there is a unique ground state configuration defined to be the path with lowest weight. 
 Let the  weight of the ground state configuration be $\w_{{\rm gsc}}$.  The relative weight of a path is $\Delta \w =  \w-\w_{{\rm gsc}}$. 
 It is the relative weight which is related to the conformal dimension of the corresponding descendant state relative to the highest-weight state (whose identification is given below).


An alternative weight prescription for RSOS paths, denoted by $w$, is presented in \cite{FLPW}. In the terminology of these references, a vertex is either scoring, with a non-zero weight $u_i$ or $v_i$ given by
 \begin{equation}\label{defuv}
 u_i = \frac12 (i-\ell_i+\ell_0) \;, \qquad v_i = \frac12(i+\ell_i-\ell_0),
\end{equation}  or non-scoring, meaning  that it does not contribute to the weight. The scoring vertices of weight $u_i$ are those where the path enters a gray band from below and the local maxima in white bands, while scoring vertices of weight $v_i$ are those where the path enters a gray band from above and the local minima that are not in gray bands. 
%
%
 
%
The scoring vertices for the three paths of Fig. \ref{fig1}-\ref{fig3} are indicated by white dots when their weight is $u_i$ and by black dots 
if it is $v_i$. The pattern of scoring vertices (their number and their position) is manifestly very sensitive to the value of $p'$.
 
That relatively few vertices contribute to the weight  $w$ is a first important simplification. The second one is that both $u_i$ and $v_i$ are always non-negative (in contrast to the expression of $\w_i$ in (\ref{wFB})). Finally, this weight prescription is absolute: there is no need to subtract the weight of the ground state with the same boundaries:
\begin{equation}
\Delta \w= \w-\w_{\text {gsc}}= w
\end{equation}

It is clear form this new weight function that
a weightless infinite tail -- mandatory for a well-defined infinite length limit in which the conformal states are recovered  -- is a zigzag within a specific gray band. Since there are $p'-1$ such bands, there are $p'-1$ possible tails for the paths. Moreover, because every path is required to terminate with a SE edge, a path must thus end at the bottom of the band, at $\ell_L=\lf rp/p'\rf$, for some $r$ within the range $1\leq r \leq p'-1$.

The proper relation between $(\ell_0,\ell_L)$ and the irreducible indices $(r,s)$, where $1\leq r\leq p'-1$ and $1\leq s\leq p-1$, is the following:
\begin{equation}\label{kacla}
\ell_0=s , \qquad \ell_L= \l\lf \frac{rp}{p'}\r \rf.
\end{equation}
Clearly, the ground-state path with $s,r$ given is the shortest path starting from $s$ that reach the $r$-th gray band and it is easily checked that it has $w=0$.

The path generating function with specified boundaries is:
\begin{equation}
X^{(p',p)}_{\ell_0, \ell_L}(q;L) = \sum_{\substack{\text{paths with fixed end}\\\text{points $\ell_0$ and $\ell_L=\ell_{L-1}-1$}\\ \text{and}\, L=\ell_0+\ell_L \, \text{mod}\, 2}}q^{w }.
\end{equation}
With this identification, $X^{(p',p)}_{\ell_0, \ell_L}(q)$ is the finitized version of the Virasoro characters $\chi^{(p',p)}_{r,s}(q)$. The full character is recovered in the limit $L\rw \y$:
\begin{equation}\label{cara}
\chi^{(p',p)}_{r,s}(q) = \lim_{L\rw \y} X^{(p',p)}_{s,\lf rp/p' \rf }(q;L). 
\end{equation}


\section{From RSOS$(p',p)$ to  B$(p',p)$ paths}

 
 We are now in position to formulate the bijection between RSOS$(p',p)$ paths to generalized Bressoud -- or B$(p',p)$ -- paths. 
 The operations defining the correspondence from RSOS$(p',p)$ to B$(p',p)$ paths are :
 \begin{itemize}
 \item Flatten  the $p'-1$ gray bands.
 
 \item  Fold the part of the strip below the first gray band $(1\leq h\leq h_1)$
 onto the region just  above it ($h_1+1\leq h\leq 2h_1$).
\item  Reset the starting height to 0.
 \end{itemize}
 
 In the flattening process, all SE and NE edges within the gray bands are mapped into H ones. It is thus crucial for the reconstruction of the RSOS path by the inverse operation that the gray bands be isolated. It is only in this way that one can recover a unique zigzag pattern in a gray band from a sequence of H edges. 
Gray bands are isolated when the condition (\ref{bou}) is satisfied.
  Our analysis is thus  restricted to those classes of models.
  Note also that the folded part  can never overlap with the second gray band.
 
The new paths are defined in the strip $0\leq x\leq L$ and $0\leq y\leq y_{{\rm max}}$ with 
\begin{equation}\label{Bup}
y_{{\rm max}}= p-p'- \l \lf \frac{p}{p'}\r \rf .
\end{equation}
 In these Bressoud-type paths, H edges are allowed at all height $y(r')$, with $1\leq r'\leq p'-1$ given by:
 \begin{equation}\label{Hedge}
y(r')= \l \lf \frac{r'p}{p'}\r \rf -\l \lf \frac{p}{p'}\r \rf -  r'+1.
\end{equation}
A path terminates at one of the height  $y(r')$, namely the one with $r'=r$.

The origin of H edges as RSOS edges within the gray bands places some restrictions on the parity of the number of successive H edges at a given height $>0$. 
If the sequence of  H edges is separated by edges of the same type, (either both NE or SE), then the number of successive H edges must be odd. It is even otherwise. In other words, if $\ell $ and $\ell'$ are the height of the vertices before and after those linked by the sequence of H edges, the number of H edges has the same parity as $|\ell-\ell'|/2$.
 
 The flattening and the folding processes make the definition of the boundary conditions at the beginning of the $B(p',p)$ path somewhat complicated in general, although these are easily read off the original RSOS path. For instance, if $s=1$ and $p>2p'-1$ (meaning that the RSOS rectangle does not have a gray band starting at $h=1$), the $B(p',p)$ path starts with a SE edge at
 $y_0=h_1-1$. Moreover, there is a constraint on the initial point $i_0$ of the first H edge, which is $i_0=h_1-1$ mod 2. Similarly, if $s=1$ and $p=2p'-1$, the first edge in the generalized Bressoud path is an H edge on the $x$ axis.

 Finally, the weight ${\hat w}$ of these new paths is computed from the sum of the $x$-position of its peaks and half the $x$-position of its half-peaks:
 \begin{equation}\label{wbre}
{\hat w} =\sum_{\text{peaks}} i +\frac12 \sum_{\text{half-peaks}}i.
\end{equation}
A half-peak is  a vertex in-between (NE,H) or (H,SE) edges.

  For a path with given boundaries $y_0= s$ and $y_L=y(r)$, the weight computed must be renormalized (by a simple subtraction) relative to the ground-state configuration pertaining to these end points:
   \begin{equation}\label{whw}
\Delta {\hat w} ={\hat w}-{\hat w}_{\text{gsc}} = w.
\end{equation}
 
 The generalized Bressoud paths corresponding to the three RSOS paths of Fig. \ref{fig1}-\ref{fig3} are illustrated in Fig. {\ref{fig4}-6. 
 

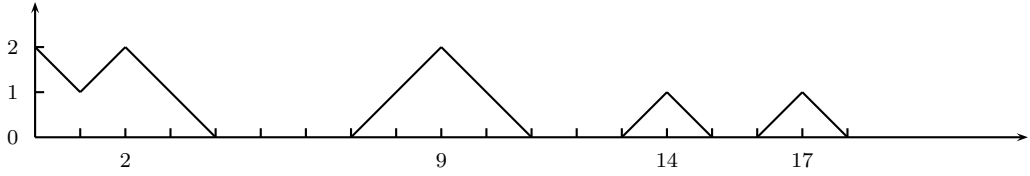
\begin{figure}[ht]
\caption{{\footnotesize The transformation of the RSOS$(2,7)$ path of Fig. \ref{fig1} into a B(2,7) path.}}
\vskip-1.4cm\label{fig4}
\begin{center}
\begin{pspicture}(3,-0.5)(10,3)

{\psset{yunit=0.6cm,xunit=0.6cm,linewidth=.8pt}


\psline{->}(0,0)(0,3)
\psline{->}(0,0)(22,0)


\psline(0,1)(0.2,1)
\psline(0,2)(0.2,2)

\psline(1,0)(1,.2)     \psline(2,0)(2,.2)        \psline(3,0)(3,.2)
\psline(4,0)(4,.2)     \psline(5,0)(5,.2)        \psline(6,0)(6,.2)
\psline(7,0)(7,.2)     \psline(8,0)(8,.2)        \psline(9,0)(9,.2)
\psline(10,0)(10,.2)     \psline(11,0)(11,.2)        \psline(12,0)(12,.2)
\psline(13,0)(13,.2)     \psline(14,0)(14,.2)        \psline(15,0)(15,.2)
\psline(16,0)(16,.2)     \psline(17,0)(17,.2)        \psline(18,0)(18,.2)

\rput(-.5,0){\scriptsize 0}
\rput(-.5,1){\scriptsize 1}
\rput(-.5,2){\scriptsize 2}

\rput(2,-0.5){\scriptsize 2}   
\rput(9,-0.5){\scriptsize 9}    
\rput(14,-0.5){\scriptsize 14} 
\rput(17,-0.5){\scriptsize 17}    



\psline(0,2)(1,1)
\psline(1,1)(2,2)
\psline(2,2)(4,0)
\psline(7,0)(9,2)
\psline(9,2)(11,0)
\psline(13,0)(14,1)
\psline(14,1)(15,0)
\psline(16,0)(17,1)
\psline(17,1)(18,0)

\psset{dotsize=2pt}\psset{dotstyle=*}
}

\end{pspicture}
\end{center}
\end{figure}

\begin{figure}[ht]
\caption{{\footnotesize The B(3,7) path corresponding to the RSOS(3,7) path of Fig. \ref{fig2}. Here, H edges allowed at $y=0,1$ but not $y=2$. The horizontal positions that are underlined indicate the positions of the half-peaks.}}
\vskip1cm
\label{fig5}
\begin{pspicture}(-0.5,-1.5)(13,1.5)

{\psset{yunit=0.6cm,xunit=0.6cm,linewidth=.8pt}


\psline{->}(0,0)(0,3)
\psline{->}(0,0)(25,0)

\psline(0,1)(0.2,1)
\psline(0,2)(0.2,2)

\psline(1,0)(1,.2)     \psline(2,0)(2,.2)        \psline(3,0)(3,.2)
\psline(4,0)(4,.2)     \psline(5,0)(5,.2)        \psline(6,0)(6,.2)
\psline(7,0)(7,.2)     \psline(8,0)(8,.2)        \psline(9,0)(9,.2)
\psline(10,0)(10,.2)     \psline(11,0)(11,.2)        \psline(12,0)(12,.2)
\psline(13,0)(13,.2)     \psline(14,0)(14,.2)        \psline(15,0)(15,.2)
\psline(16,0)(16,.2)     \psline(17,0)(17,.2)        \psline(18,0)(18,.2)
\psline(19,0)(19,.2)    \psline(20,0)(20,.2)        \psline(21,0)(21,.2)
\psline(22,0)(22,.2)     \psline(23,0)(23,.2)        \psline(24,0)(24,.2)


\rput(-.5,0){\scriptsize 0}
\rput(-.5,1){\scriptsize 1}
\rput(-.5,2){\scriptsize 2}

\rput(2,-0.5){\scriptsize 2}
\rput(5,-0.5){\scriptsize 5}    \rput(7,-0.5){\scriptsize {\underline 7}}
\rput(9,-0.5){\scriptsize 9}    \rput(11,-0.5){\scriptsize \underline {11}}
\rput(13,-0.5){\scriptsize \underline {13}}    \rput(15,-0.5){\scriptsize \underline {15}}
  \rput(19,-0.5){\scriptsize 19}
\rput(21,-0.5){\scriptsize 21}    \rput(23,-0.5){\scriptsize 23}



\psline(0,1)(1,0)
\psline(1,0)(2,1)
\psline(2,1)(3,0)
\psline(3,0)(4,0)
\psline(4,0)(5,1)
\psline(5,1)(6,0)
\psline(6,0)(7,1)
\psline(7,1)(8,1)
\psline(8,1)(9,2)
\psline(9,2)(10,1)
\psline(10,1)(11,1)
\psline(11,1)(12,0)
\psline(12,0)(13,1)
\psline(13,1)(15,1)
\psline(15,1)(16,0)
\psline(16,0)(18,0)
\psline(18,0)(19,1)
\psline(19,1)(20,0)
\psline(20,0)(21,1)
\psline(21,1)(22,0)
\psline(22,0)(23,1)
\psline(23,1)(24,0)

\psset{dotsize=2pt}\psset{dotstyle=*}
}

\end{pspicture}
\end{figure}
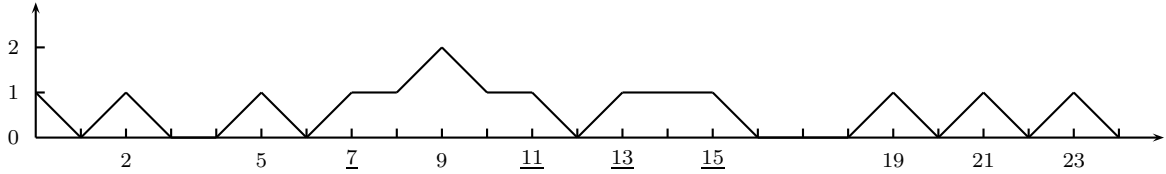

\begin{figure}[ht]\label{fig6a}
\caption{{\footnotesize The B(4,7) path corresponding to the RSOS(4,7) path of Fig. \ref{fig3}. Here, H edges are allowed at all values of $y\leq 2$. The horizontal positions that are underlined indicated the positions of the half-peaks.}}
\begin{center}
\begin{pspicture}(1.5,-0.5)(13,2)

{\psset{yunit=0.6cm,xunit=0.6cm,linewidth=.8pt}


\psline{->}(0,0)(0,3)
\psline{->}(0,0)(25.5,0)

\psline(0,1)(0.2,1)
\psline(0,2)(0.2,2)

\psline(1,0)(1,.2)     \psline(2,0)(2,.2)        \psline(3,0)(3,.2)
\psline(4,0)(4,.2)     \psline(5,0)(5,.2)        \psline(6,0)(6,.2)
\psline(7,0)(7,.2)     \psline(8,0)(8,.2)        \psline(9,0)(9,.2)
\psline(10,0)(10,.2)     \psline(11,0)(11,.2)        \psline(12,0)(12,.2)
\psline(13,0)(13,.2)     \psline(14,0)(14,.2)        \psline(15,0)(15,.2)
\psline(16,0)(16,.2)     \psline(17,0)(17,.2)        \psline(18,0)(18,.2)
\psline(19,0)(19,.2)     \psline(20,0)(20,.2)        \psline(21,0)(21,.2)
\psline(22,0)(22,.2)    \psline(23,0)(23,.2)    \psline(24,0)(24,.2)    
\psline(25,0)(25,.2)    


\rput(-.5,0){\scriptsize 0}
\rput(-.5,1){\scriptsize 1}
\rput(-.5,2){\scriptsize 2}

\rput(4,-0.5){\scriptsize \underline 4}    \rput(8,-0.5){\scriptsize  \underline 8}
\rput(10,-0.5){\scriptsize  \underline {10}}
\rput(14,-0.5){\scriptsize 14}    \rput(16,-0.5){\scriptsize  \underline {16}}
\rput(18,-0.5){\scriptsize  \underline {18}}
\rput(24,-0.5){\scriptsize  \underline {24}}



\psline(3,0)(4,1)
\psline(4,1)(7,1)
\psline(7,1)(8,2)
\psline(8,2)(10,2)
\psline(10,2)(11,1)
\psline(11,1)(13,1)
\psline(13,1)(14,2)
\psline(14,2)(15,1)
\psline(15,1)(16,1)
\psline(16,1)(17,0)
\psline(17,0)(18,1)
\psline(18,1)(24,1)
\psline(24,1)(25,0)


\psset{dotsize=2pt}\psset{dotstyle=*}

}

\end{pspicture}
\end{center}
\end{figure}


The bijective nature of this correspondence is  clear. On the one hand, the transformation of a RSOS to a generalized Bressoud path is manifestly unique. On the other hand, the reverse procedure is also unambiguous.
With both $L$ and the nature  of the last edge (SE) fixed, the RSOS path is uniquely recovered from a B path, starting from the end (since the folding does not induce an ambiguity on the end point). 
In this reverse operation, each line where there are allowed values for the H edges is transformed into a gray band: H edges become zigzag patterns within gray bands. 

 The crucial point is the demonstration of the  weight preserving character of the bijection, namely, the proof of (\ref{whw}). To present the key point without the complications induced by the boundary effects, let us restrict to the case $s=r=1$, for which ${\hat w}_{\text{gsc}}=0$. What has to be shown then is that $w={\hat w}$.
This is demonstrated in terms of a pairing of the scoring vertices of different types, from right to left and in a nested way (meaning that adjacent unlike scoring vertices are paired first, as clarified in the examples below). If the scoring vertices have height $\ell_i> h_1$, the pairing starts (at the right) with a $\bullet$ vertex; this is called a type-1 pairing. A type-2 pairing occurs when the two vertices are below the first gray band; in that case, the pairing  starts with a $\circ$-type vertex.
With the specified boundaries, all scoring vertices come in pair except for an isolated $\circ$ at the beginning of the path (if $p>2p'-1$), whose weight is 0 (and such unpaired vertex is absent when $p=2p'-1$).

For the path of Fig. \ref{fig1}, the pairing is as follows: $\circ (\bullet \circ) (\circ \bullet)(\circ \bullet)(\bullet \circ)$. All pairings are between nearest neighbors. For the path of Fig. \ref{fig2}, the pairing is $\circ (\bullet \circ) (\circ \bullet)(\circ (\circ \bullet)\bullet)(\circ \bullet)(\circ \bullet)(\circ \bullet)(\circ \bullet)$. In this case, there is a nested pairing: the  $\circ$ vertex at 7 is paired with the  $\bullet$ vertex at 12 and these are separated by a pair $(\circ \bullet)$.  A non-nearest-neighbor pairing occurs when the path  in-between the two scoring vertices crosses one or more gray bands.
For the path  in Fig. {\ref{fig3}, 
the pairings are $(\circ(\circ \bullet) (\circ \bullet) \bullet)(\circ \bullet)$: the $\circ$ vertex  at 4 is paired with the $\bullet$ one at 17.

If a pair is not separated by a zigzag pattern in a gray band, it is mapped (via the bijection) to a peak centered at the position $i$ of the left-most scoring vertex of the pair. If there is an in-between zigzag pattern in a gray band of length $2\beta$ (which is  possible only for type-1 pairing), then the pair is replaced by 
a pair of half-peaks, one at $i$ and the other at $i+2\beta$. The different possibilities are illustrated in Fig. \ref{fig7}.


The correspondence just indicated is supported by the verification of the
identity  $w={\hat w}$. 
Consider a pairing of the first type, with the vertices  $(\circ,\bullet)$ at respective position $(i,\ell_i)$ and $(i+a,\ell_i-a)$, so that 
\begin{equation}
u_\circ=\frac{i-\ell_i+1}{2},\qquad v_\bullet= \frac{i+a+(\ell_i-a-1)}{2}\qquad \Rw \qquad w_{(\circ,\bullet)}=u_\circ+v_\bullet=i,
\end{equation}
which is the weight of the Bressoud peak at position $i$ (cf. Fig \ref{fig7}a).
In the more general case where the two scoring vertices are separated by a zig-zag pattern of length $2\beta$, one has
\begin{equation}
u_\circ=\frac{i-\ell_i+1}{2},\qquad v_\bullet= \frac{i+a+2\beta+(\ell_i-a-1)}{2}\qquad \Rw \qquad w_{(\circ,\bullet)}=u_\circ+v_\bullet=\frac{i}2+\frac{i+2\beta}2.
\end{equation}
Here the total weight $i+\beta$ is written in the form of two contributing half-peaks, at $i$ and $i+2\beta$ (cf. Fig. \ref{fig7}b).
In both cases, the pair of scoring vertices is found to describe a peak of height $a$, which is flatten if $\beta>0$.  The computation for a case where there are in-between pairs of scoring vertices is similar. In such a case, the path leaves the gray band from above, instead of simply zigzagging within it; there will then necessarily be at least one pair of unlike scoring vertices in the portion of the path above the gray band under consideration. The result of the computation is the same as above, with $2\beta$ being defined more precisely as the distance between the gray band entry point (the position of the $\circ$ vertex) and its exit point in the downward direction.

When the pairing is of the form $(\bullet,\circ)$, with the scoring vertices at respective position $(i,\ell_i)$ and $(i+a,\ell_i+a)$, with $\ell_i+a\leq h_1$ (since for that type of pairing the two vertices must lie below the first gray band),  $w_{(\bullet,\circ)}= i$; the peak in the B path is centered at $i$ (cf. Fig. \ref{fig7}c).


 This analysis shows that pairs of scoring vertices can be put in correspondence with peaks or pairs of half-peaks and that this correspondence is weight preserving.
This completes the proof of the identity $w={\hat w}$ for $s=r=1$.


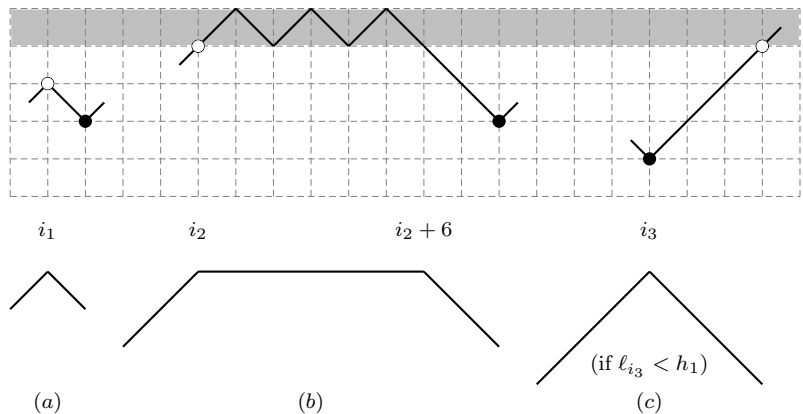
\begin{figure}[ht]
\caption{{\footnotesize The different types of pairing of two scoring vertices that make a peak, or two paired half-peaks, in the corresponding generalized Bressoud path. The first two cases describe pairings of type 1, that is of the form $(\circ,\bullet)$. In (a) the two scoring vertices are adjacent local extrema. The two vertices `fuse' in the Bressoud path into a simple peak. The height of this peak is given by the height difference between the two scoring vertices, here 1. The position of the peak is that of the $\circ$ vertex. In (b), the two vertices are separated by a gray-band zigzag pattern of length 6. The corresponding half-peaks are separated by 6 H edges. The positions of the half-peaks are at the extremities of the zigzag pattern, namely at the position of the $\circ$ and that of the $\bullet $ minus the height difference, which is 2 here. This  height difference is the  height of the resulting flatten peak. The case (c) describes a situation that is possible only if the two scoring vertices lie below the first gray band (with $r'=1$). They both fuse into a peak (due to the folding process) and the peak position is that of the $\bullet$ vertex.  Its  height is 3, the  height difference of the two scoring vertices.}}
\vskip-1.4cm
\label{fig7}
\begin{center}
\begin{pspicture}(-2,-3)(13.0,4.5)



{
\psset{linestyle=none}
\psset{fillstyle=solid}
\psset{fillcolor=lightgray}
\psframe(0.5,1.5)(11, 2.0)
}




{
\psset{linewidth=0.25pt,linestyle=dashed, dash=2.5pt 1.5pt,linecolor=gray}

\psline{-}(0.5,-0.5)(11,-0.5)
\psline{-}(0.5,0.0)(11,0.0)
\psline{-}(0.5,0.5)(11,0.5)
\psline{-}(0.5,1.0)(11,1.0)
\psline{-}(0.5,1.5)(11,1.5)
\psline{-}(0.5,2.0)(11,2.0)

\psline{-}(0.5,-0.5)(.5,2.0)
\psline{-}(1.0,-0.5)(1.0,2.0) \psline{-}(1.5,-0.5)(1.5,2.0) 
\psline{-}(2.0,-0.5)(2.0,2.0) \psline{-}(2.5,-0.5)(2.5,2.0) \psline{-}(3.0,-0.5)(3.0,2.0) 
\psline{-}(3.5,-0.5)(3.5,2.0) \psline{-}(4.0,-0.5)(4.0,2.0) \psline{-}(4.5,-0.5)(4.5,2.0) 
\psline{-}(5.0,-0.5)(5.0,2.0) \psline{-}(5.5,-0.5)(5.5,2.0) \psline{-}(6.0,-0.5)(6.0,2.0) 
\psline{-}(6.5,-0.5)(6.5,2.0) \psline{-}(7.0,-0.5)(7.0,2.0) \psline{-}(7.5,-0.5)(7.5,2.0) 
\psline{-}(8.0,-0.5)(8.0,2.0) \psline{-}(8.5,-0.5)(8.5,2.0) \psline{-}(9.0,-0.5)(9.0,2.0) 
\psline{-}(9.5,-0.5)(9.5,2.0) \psline{-}(10.0,-0.5)(10.0,2.0) \psline{-}(10.5,-0.5)(10.5,2.0) 
\psline{-}(11.0,-0.5)(11.0,2.0)
}




\psset{linestyle=solid}


\psline{-}(0.75,.75)(1,1.0)\psline{-}(1,1.0)(1.5,0.5)\psline{-}(1.5,0.5)(1.75,.75)

\rput(1,-0.95){\scriptsize {$i_1$}}    
\psline{-}(0.5,-2)(1,-1.5)\psline{-}(1,-1.5)(1.5,-2)
\psline{-}(2.75,1.25)(3.0,1.5)
\psline{-}(3.0,1.5)(3.5,2.0) \psline{-}(3.5,2.0)(4.0,1.5)
\psline{-}(4.0,1.5)(4.5,2.0) \psline{-}(4.5,2.0)(5.0,1.5)
\psline{-}(5.0,1.5)(5.5,2.0) \psline{-}(5.5,2.0)(6.0,1.5)
\psline{-}(6.0,1.5)(6.5,1.0) \psline{-}(6.5,1.0)(7.0,0.5)
\psline{-}(7.0,0.5)(7.25,0.75)

\rput(3,-0.95){\scriptsize {$i_2$}}
\rput(6,-0.95){\scriptsize {$i_2+6$}}        
\psline{-}(2.0,-2.5)(3,-1.5)\psline{-}(3,-1.5)(6,-1.5)\psline{-}(6,-1.5)(7,-2.5)

\psline{-}(8.75,0.25)(9,0.0)
\psline{-}(9,0.0)(10.5,1.5)\psline{-}(10.5,1.5)(10.75,1.75)

\rput(1,-3.25){\scriptsize {$(a)$}}    
\rput(4.5,-3.25){\scriptsize {$(b)$}}    
\rput(9,-3.25){\scriptsize {$(c)$}}    
\rput(9,-0.95){\scriptsize {$i_3$}}    
\psline{-}(7.5,-3)(9,-1.5)\psline{-}(9,-1.5)(10.5,-3)
\rput(9,-2.75){\scriptsize {($\text{if}\; \ell_{i_3}<h_1$)}}



\psset{fillcolor=white}
\psset{dotsize=5pt}\psset{dotstyle=o}
\psdots(1,1.)(3,1.5)(10.5,1.5)
\psset{dotsize=5pt}\psset{dotstyle=*}
\psdots(1.5,0.5)(7.0,0.5)(9,0)
\end{pspicture}
\end{center}
\end{figure}


 

 \section{Sample $B(p',p)$ particle spectra and generating functions}

 We identify here the particle spectrum of some classes of B paths and outline the derivation of their generating function along the lines of \cite{OleJS} (see also \cite{JSTAT,JMsusy}). The resulting fermionic forms  are given for the vacuum module and in the limit $L\rw \y$. The detail of these derivations  will be presented elsewhere, together with the analysis of the additional modules.

 \subsection{B$(2,2k+1)$ paths}


As already said, a RSOS$(2,2k+1)$ path is defined within a rectangle having a single gray band in its center. In the transformation, this band is flatten out, the part of the strip below the gray band is folded
 onto the one above it, and the height $h=k+1$ is reset to $y=0$. 
 By construction,  this  new path has peaks with height $\leq k-1$.  These peaks are either at the positions of the peaks in the portions of the RSOS$(2,2k+1)$ path which lie  above the gray band or at the positions of the valleys below it. In addition, the B$(2,2k+1)$ paths might contain possible H edges on the horizontal axis: these are the edges that originally lie within the gray band.  
 Clearly, the resulting  paths have all the characteristics of Bressoud paths.\footnote{We noticed afterwards -- and with insight -- that this bijection (without  the folding operation) is also pointed out in \cite{WelPro} for the  special case of the $\M(2,5)$ model.}
 
Note that for this special case,  a weight-preserving bijection is known to  exist (combining the results of \cite{ABBBFV} and \cite{BreL}). Once the relationship between the two types of paths is unraveled, it follows that the correspondence must be weight preserving. This provides an alternative proof of the identity $w={\hat w}$ for that case (and which applies here to all values of $s$).

 Now, the fermi-gas analysis of the Bressoud paths is straightforward.\footnote{It is spelled out in \cite{W97} in terms of a different type of paths. However, the  analysis of \cite{OleJS} is readily adapted to Bressoud paths. Moreover, although the  generating function in \cite{BreL} is obtained by an inductive argument (which is thus a verification proof rather than a constructive one), it contains the key features for the constructive fermi-gas derivation.} The particles are the peaks of height $1\leq y\leq k-1$. We let the height of an isolated peak to be its charge.
Let $m_j$ be the number of peaks of charge $m_j$ in the path. With a fixed set of values $\{m_j\}$, the first step is to determine the configuration of minimal weight. Here it corresponds to the case where all the peaks are ordered in increasing value of the charge. The corresponding weight is  $mBm +Cm$ in the expression below. Then, we determine the possible displacements of each type of peaks subject to two rules: particles with the same charge do not penetrate and particles of different charge can penetrate each other as long as they do not generate higher charged particles. In the present case, the presence of other types of particle does not affect the displacements of a particle of a given charge and each displacement of 1 step increases the weight by 1. 
To each type of particle, there is thus a factor $(q)_{m_j}^{-1}$.
Then one sums over all values of $m_j\geq 0$.  
 This leads directly to the fermionic character \cite{FNO}:
\begin{equation}
\chi^{(2,2k+1)}_{1,1}(q)= \sum_{m_1,\cdots,m_{k-1}} \frac{
q^{mBm+Cm} } { (q)_{m_1}\cdots (q)_{m_{k-1}} },
\end{equation}
where 
\begin{equation}
(q)_m= \prod_{i=1}^m (1-q^i),\qquad 
B_{i,j}= \text{min}\,  (i,j) \quad  \text{and}\quad  C_j=j.
\end{equation}
\subsection{B$(p',kp'-1)$ paths}

We now consider a natural  extension of the previous class of models, the minimal models $\M(p',kp'-1)$, with $k>1$. When $p=kp'-1$, the RSOS defining rectangle has a completely regular band structure: there are $k-2$ white bands at the borders of the rectangle (below the first gray band and above the last gray band) and two consecutive gray bands are separated by $k-1$ white bands. This class of models incorporates the limiting case $p=2p'-1$ and Fig. \ref{fig3} illustrates its band structure when $p'=4$.

The corresponding B$(p',kp'-1)$ paths can be decomposed into a sequence of $p'+k-4$ particles labeled by $j=1,\cdots, p'+k-4$ and these are naturally separated into two sets: 
\begin{equation}\label{sets}
I=\{1,\cdots, k-2\}\qquad \text{ and} \qquad I'=\{k-1,\cdots , p'-k-4\}.
\end{equation}
 The first set of $k-2$ particles is defined as in the B$(2,2k-1)$ case: these are triangles of height $j$, with $j\in I$. For the other particles, it is convenient to define  $j'=j-k+2$, so that $1\leq j' \leq p'-2$. The corresponding particles 
are represented by deformed symmetric triangles of height $j'(k-1)$ and whose  left-side has $j'$ sequences  $k-1$ NE edges all separated by an H edge. The top is a peak at height $j'(k-1)$, a height where H edges are allowed. It is thus convenient to view the top peak as being composed of two half-peaks. In addition to these, the particle labeled by $j'$ contains  $2(j'-1)$ other half-peaks due the the presence of internal H edges. See for instance Fig. \ref{fig9} where the particles for $p'=5,\, k=3$ are pictured; in that case, there are two particles of each type.

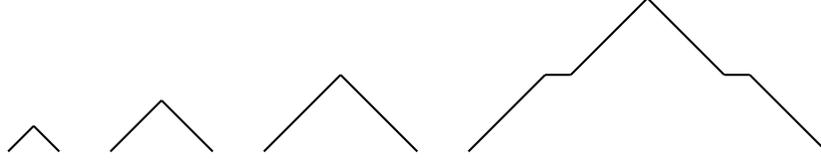
\begin{figure}[ht]
\caption{{\footnotesize The four basic particles for the B$(5,14)$ path (of the class $(p',kp'-1)$ with $p'=5$ and $k=3$). The first two are labeled by the set $I$ and the last two by the set $I'$ in (\ref{sets}). H edges are allowed at height 0, 3 and 6.}}
\vskip-1.cm
\label{fig9}
\begin{center}
\begin{pspicture}(1,0)(10,2.5)

{\psset{yunit=0.34cm,xunit=0.34cm,linewidth=.8pt}


{\psset{linestyle=none}
\psset{fillstyle=solid}
\psset{fillcolor=yellow!30!white}
}

\psline(1,0)(2,1)\psline(2,1)(3,0)
\psline(5,0)(7,2)\psline(7,2)(9,0)
\psline(11,0)(14,3)\psline(14,3)(17,0)

\psline(19,0)(22,3)\psline(22,3)(23,3)
\psline(23,3)(26,6)\psline(26,6)(29,3)
\psline(29,3)(30,3)\psline(30,3)(33,0)
}

\end{pspicture}
\end{center}
\end{figure}

\def\1{{\bar 1}}
In the minimal-weight configuration (for a fixed particle content $\{m_j\}$), the particles are ordered  by increasing values of $j\in I\cup I'$.

Each type of particles has a different dynamics. For those labeled by $j\in I$, 
the displacements are free (the presence of other particles of either type does not induce constraints) and they are described by the factor $(q)_{m_j}^{-1}$.  This is as for the B$(2,2k-1)$  particles. The dynamics of the other type of particles is best regarded in terms of that of the half-peaks, taken collectively. Every half-peak is allowed to be displaced by 2 units, with a weight difference of 1. Let $p_{k-2}$ be the number of  half-peaks, where $p_j$ is defined as follows:
\begin{equation}\label{defpj}
p_j= 2m_{j+1}+4m_{j+2}+\cdots +2(p'+k-3-j)m_{p'+k-4} \qquad \text{for}\quad j+1\in I'.
\end{equation}
The displacements of all half-peaks  is governed by the factor $(q)_{p_{k-2}}^{-1}$. However, this does not capture all possible configurations. 
Given a configuration with a specific set of positions  for the various half-peaks, it is still possible to move half-peaks  in-between the configuration under consideration -- maintaining the two extremal (the first and the last) half-peaks fixed -- such that each basic displacement increases the weight by 1. A basic displacement  typically involves the opposite displacements of two half-peaks. The resulting configurations are most simply described in terms of  the different penetration patterns of smaller particles (but still within the set $I'$) into larger ones.  All these possibilities are taken into account by the product of $q$-binomials in the expression below.

For instance, with the edges NE, SE and H represented respectively by $1,\1$ and 0, the three configurations with $m_3=m_4=1$,
regarded e.g. as  B(5,14) paths (cf. Fig. \ref{fig9}), are :
\begin{align}
& 1 1 1  \1\1\1 111 0 111 \1\1\1 0 \1\1\1 \nonumber\\
& 1 1 1 0 111 \1\1\1  111 \1\1\1 0 \1\1\1 \nonumber\\
& 1 1 1 0 111 \1\1\1 0 \1\1\1 111  \1\1\1 .
\end{align}
A peak (= two half-peaks) is the vertex in-between $1\1$ and a half-peak is either in-between $10$ or $ 0\1 $.  A configuration is obtained from the previous one by displacing one half-peak  forward by 4 units and another one backward by 2 units, for a weight difference of 1. The results are taken into account by the $q$-deformation of the binomial factor $\binom{3}{1}$ defined below.

The vacuum character resulting from the analysis just sketched reads 
\begin{equation}
\chi_{1,1}^{(p',p'k-1)}(q)= \sum_{m_1,\cdots,m_{p'+k-4}} \frac{ q^{mBm+Cm}}{(q)_{m_1}\cdots (q)_{m_{k-2}} (q)_{p_{k-2}}}\;  \prod_{i=k-1}^{p'+k-5}
\begin{bmatrix} m_{i}+p_i\\ m_i
\end{bmatrix},
\end{equation}
where $p_j$ is defined above. $B_{i,j}$ is the symmetric matrix whose entries, with $i\leq j$, are 
\begin{equation}\label{defBC}
B_{i,j}= \begin{cases} &
i\quad  \text{for}\quad i,j\in I\\ &
   i j'\quad  \text{for}\quad i\in I , \, j\in I'\\ &
   (i'k-1)j'\quad \text{for}\quad i,j\in I',
\end{cases}  \qquad \text{and}\qquad C_{j}= \begin{cases} &
 j \quad \text{for}\quad j\in I\\ &
  (k-1)j' \quad \text{for}\quad j\in I'.
\end{cases}
\end{equation}
(Recall that  $j'=j-k+2$.)
Finally, the $q$-binomial is defined as
\begin{equation}
\begin{bmatrix}
a\\ b\end{bmatrix} =  \begin{cases} & \frac{(q)_a}{(q)_{a-b}(q)_b}  \quad \text{if}\quad 0\leq b\leq a,\\ &\qquad 0\qquad\qquad\text{otherwise}.
\end{cases}
\end{equation}
We stress that this fermionic-character expression is not new \cite{FLPW} but it is witten there in terms of the variables $p_j$ (relabeled $p_{j+1}$) instead of $m_j$ for $j\in I'$ 
(see also \cite{FQ} Corr. 1.6, but pertaining to a different module, which entails minor modifications). 
However, its dressing with a particle interpretation is new.

\subsection{B$(p',kp'+1)$ paths}

To exemplify the non-triviality of the particle decomposition, we present the results for a related  sequence, namely, $p=kp'+1$, with $k>1$. The band structure still follows a simple pattern: there are now $k-1$ white bands below the first gray band and two consecutive gray bands are again separated by $k-1$ white bands.
The generalized Bressoud  paths are described in terms of $p'+k-3$ particles, naturally divided into two sets, labeled by $I$ and $I'$ defined as before in (\ref{sets}) but with $k\rw k+1$. The first $k-1$ particles again correspond to  triangles of height $j$, with $1\leq j\leq k-1$. The other particles, with $j=k-1+j'$, correspond to deformed and flatten symmetric triangles of height $j'(k-1)$ and whose  left-side has $j'$ sequences of $k-1$ NE edges, all separated by an H edge. In addition, these particles are topped by two H edges  and they start and end with an H edge. See Fig. \ref{fig10} for the representation of the particles of the B$(4,13)$ paths.

\begin{figure}[ht]
\caption{{\footnotesize The four basic particles for the B$(4,13)$ paths (of the class $(p',kp'+1)$ with $p'=4$ and $k=2$). 
The second and the third particle both have height 2 but they differ in that the third one can be opened at its top but not the second one. Here H  are  allowed at height 0, 2 and 4.}}
\vskip-.5cm\label{fig10}
\begin{center}
\begin{pspicture}(1,-0.5)(10,2.)

{\psset{yunit=0.34cm,xunit=0.34cm,linewidth=.8pt}


{\psset{linestyle=none}
\psset{fillstyle=solid}
\psset{fillcolor=yellow!30!white}
}

\psline(1,0)(2,1)\psline(2,1)(3,0)
\psline(5,0)(7,2)\psline(7,2)(9,0)
\psline(11,0)(12,0)\psline(12,0)(14,2)
\psline(14,2)(16,2)\psline(16,2)(18,0)
\psline(18,0)(19,0)

\psline(21,0)(22,0)\psline(22,0)(24,2)
\psline(24,2)(25,2)\psline(25,2)(27,4)
\psline(27,4)(29,4)\psline(29,4)(31,2)
\psline(31,2)(32,2)\psline(32,2)(34,0)
\psline(34,0)(35,0)
}

\end{pspicture}
\end{center}
\end{figure}
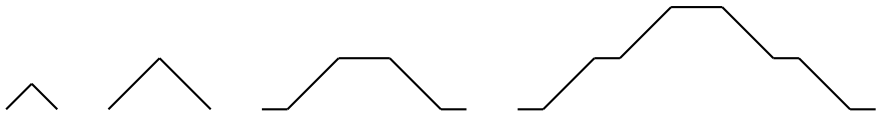
\vskip-0.5cm

For the minimal-weight configuration,
the particles are ordered  by increasing values of $j\in I $ and then by decreasing values of $j\in I'$.
The rest of the analysis is similar to the one sketched in the previous case and results into
\begin{equation}
\chi_{1,1}^{(p',p'k+1)}(q)= \sum_{m_1,\cdots,m_{p'+k-3}} \frac{ q^{mBm+Cm}}{(q)_{m_1}\cdots (q)_{m_{k-1}} (q)_{p_{k-1}}}\;  \prod_{i=k}^{p'+k-4}
\begin{bmatrix} m_{i}+p_i\\ m_i
\end{bmatrix},
\end{equation}
where $p_j$ is defined as before but with $k\rw k+1$.
The form of the  symmetric matrix $B_{i,j}$ 
is as in (\ref{defBC}) except for the third condition which reads : $B_{i,j}=i'(j'k+1)$ if $i\leq j$ and  $i,j\in I'$.
Finally, $C$ is again given by eq. (\ref{defBC}).
Again this expression is known \cite{FLPW}
(see also the second reference in \cite{KKMMa} Sect. 4.4 and cf. \cite{FQ} Corr. 1.8), 
 but its particle interpretation is new.

\section{Particles vs kinks and breathers}

In the previous section, we have unraveled the particle content of RSOS$(p',kp'\pm 1)$ paths.
We expect the gross features of this spectrum to be generic to models with $p\geq 2p'-1$. 
These features are the following. 
There are basically two types of particles. Type-$a$ particles  are those which can live in the white bands of the RSOS path that are below the first gray band (and they appear there as valleys). This defining domain is not exclusive (they can be seen elsewhere) but it is a precise characterization: the other particles cannot live there.  The number of type-$a$ particles is thus $h_1-1=\lf p/p'\rf-1$. In the corresponding Bressoud path, they are represented by triangles of height $1\leq y\leq \lf p/p'\rf-1$; these particles cannot be deformed by the insertion of H edges.

Type-$b$ particles, on the other hand, are triangles that link  two gray bands in the RSOS path. Each such particle is thus built up from sequences of NE and SE edges interpolating between two adjacent gray bands. As there are $p'-1$ gray bands, there are thus $p'-2$ such interpolating segments.
 Since the gray bands  represent the possible RSOS vacua, type-$b$ particle constituents can be viewed as fundamental kinks or antikinks interpolating between adjacent vacua. Particles of type-$b$ are thus composed of $j'$ kinks and $j'$ antikinks, with $1\leq j'\leq p'-2$. In the Bressoud path context, a kink (antikink) interpolates between two successive heights where H edges are allowed and, in a fundamental particle, two adjacent kinks (antikinks) are separated by an H edge.

The characteristic  property of type-$b$ particles in B paths is that  they can be  deformed by the introduction of an even number of H edges at every height where such edges are allowed (and in particular, at their top); in other words, after any kink or antikink constituent segment, pairs of H edges can be  introduced.
 This means that the constituent kinks and antikinks can be separated at will: they do not form bound states.  
A kink (antikink) component of a type-$b$ particle actually corresponds to a half-peak toward the right (left). In contrast, particles of type $a$ appear to describe genuine bound states, i.e.,  two confined half-peaks. 

As mentioned in the introduction, the RSOS$(p',p)$ model in regime III provides an integrable off-critical (lattice) description of the  $\phi_{1,3}$-perturbed minimal model $\M(p',p)$. Its scaling limit is the restricted sine-Gordon equation at coupling $\beta^2/8\pi=p'/p$ \cite{LeC}. This model has $p'-1$ vacua, hence $p'-2$ kinks (and as many antikinks). Moreover, there are $\lf p/p'\rf-1$ bound states or breathers (see e.g. \cite{ZZ}). This match the above counting: particles of type $a$ are similar to breathers and those of type $b$ are like composite of $j'$  kinks and antikinks, or equivalently, pairs of kink-antikink of topological charge $j'$.\footnote{Note that the perfect pairing between the number of kinks  and antikinks encountered in most of our examples is a special feature of the condition $s=r$ (=1 for the vacuum module). For paths with $s\not=r$, there will be an excess of kinks or antikinks, a neat evidence that they are not confined, even loosely. 
But note that these are the pairs of kink-antikink of charge $j'$ that enter in the particle description of the fermionic forms, the excess being treated as a boundary effect.}

\section{Conclusion}

We have presented a bijection between the RSOS paths and new generalized Bressoud paths for the class of models with $p\geq 2p'-1$. In those cases, we have essentially provided a third weight function for the RSOS paths. The genuine advantage of this transformation of the RSOS paths to Bressoud-type ones, beyond the simplicity of the resulting weight expression, 
is that it marks out the way for a particle interpretation of the RSOS paths, which in turn should make
possible the construction of their generating function (i.e., the character) by direct fermi-gas-type method. This has been illustrated here for the $\M(p',kp'\pm1)$ minimal models and remains to be worked out in full generality.

The particle spectrum that emerges naturally from the Bressoud path formulation  -- at least in its gross features abstracted from the cases studied explicitly --- has been interpreted as the conformal limit of the   solitons and breathers spectrum of the restricted sine-Gordon model at $\beta^2/8\pi= p'/p.$
This matching provides neat support to the idea that fermionic forms capture the physics of a particular integrable perturbation and that its underlying quasi-particles 
should correspond to the massless limit of the particles  of the off-critical theory (see e.g., \cite {KKMMa}).

The present analysis is restricted to the cases $p\geq 2p'-1$. From the finitized characters of the models with $p>2p'$, the finitized characters of those  with $p<2p'$ can be obtained by duality \cite{BMlmp,FLPW}. Alternatively, 
for $p<2p'$, there is a dual version of the transformation of the RSOS paths to the generalized Bressoud ones, but with a somewhat more complicated weight function. This will be considered elsewhere. But note that for those cases, there are no breather in the spectrum since the bordering bands of the RSOS defining rectangle are always gray (or equivalently, all white bands are isolated).

%

\vskip0.3cm
\noindent {\bf ACKNOWLEDGMENTS}

PM acknowledge useful discussions with A. LeClair and  T. Welsh.  We also thanks T Welsh for his critical comments on the article.
Part of this work was done during the workshop {\it Low-dimensional Quantum Field Theories and Applications} at   
 the Galileo Galilei Institute for Theoretical Physics. PM thanks the GGI  for its hospitality and the INFN for partial support during his stay.  This  work is supported  by NSERC.

\end{document}